\documentclass[aps,prb,twocolumn,floatfix,superscriptaddress]{revtex4-1}
\usepackage{graphicx,txfonts,bm}
\usepackage[colorlinks,citecolor=magenta,linkcolor=blue]{hyperref}

\begin{document}


\title{Reexaming the ground state and magnetic properties of curium dioxide}


\author{Li Huang}
\email{lihuang.dmft@gmail.com}
\affiliation{Science and Technology on Surface Physics and Chemistry Laboratory, P.O. Box 9-35, Jiangyou 621908, China}

\author{Ruofan Chen}
\affiliation{Science and Technology on Surface Physics and Chemistry Laboratory, P.O. Box 9-35, Jiangyou 621908, China}

\author{Haiyan Lu}
\affiliation{Science and Technology on Surface Physics and Chemistry Laboratory, P.O. Box 9-35, Jiangyou 621908, China}

\date{\today}


\begin{abstract}
The ground state electronic structure and magnetic behaviors of curium dioxide (CmO$_{2}$) are controversial. In general, the formal valence of Cm ions in CmO$_{2}$ should be tetravalent. It implies a $5f^{6.0}$ electronic configuration and a non-magnetic ground state. However, it is in sharp contrast with the large magnetic moment measured by painstaking experiments. In order to clarify this contradiction, we tried to study the ground state electronic structure of CmO$_{2}$ by means of a combination of density functional theory and dynamical mean-field theory. We find that CmO$_{2}$ is a wide-gap charge transfer insulator with strong 5$f$ valence state fluctuation. It belongs to a mixed-valence compound indeed. The predominant electronic configurations for Cm ions are $5f^{6.0}$ and $5f^{7.0}$. The resulting magnetic moment agrees quite well with the experimental value. Therefore, the magnetic puzzle in CmO$_{2}$ can be appropriately explained by the mixed-valence scenario.
\end{abstract}


\maketitle


\section{introduction\label{sec:intro}}

Actinide-based materials manifest complex and fascinating magnetic behaviors~\cite{RevModPhys.81.235}. To recognize them and understand their underlying mechanisms are always hot topics in condensed matter science. In the early actinides (such as U, Np, and Pu), itinerant 5$f$ electrons result in significant band dispersions and large band widths. Hence the ratio between exchange interaction and 5$f$ band width does not meet the requirement of Stoner criterion~\cite{PhysRevLett.71.3214}, so magnetic ordering is absent in these actinides~\cite{PhysRevB.72.054416}. As for the late actinides (such as Am, Cm, Bk, and beyond), the picture is conspicuously different. The 5$f$ electrons in the late actinides should become localized. They are capable of spin polarization and realizing some kinds of magnetic ordering states. However, no macroscopic moment has been observed experimentally in Am~\cite{PhysRevLett.114.097203}. This is because its orbital and spin moments are equal but with opposite signs ($\mu_L = -\mu_S$). As a consequence, Cm comes to be the first actinide element that exhibits magnetic ordering under ambient condition~\cite{Heathman110}. It has an antiferromagnetic ground state with large ordered moment $\mu_{\text{eff}} \approx 7.58~\mu_{\text{B}}$~\cite{KANELLAKOPULOS1975713,PhysRevB.99.224419}.

Now let us focus on the actinide dioxides, $An$O$_{2}$, which crystallize in a cubic CaF$_{2}$-like structure. They usually show complicated magnetic ordering phases and thus garner much attention~\cite{wen:2013}. In the early members of the actinide dioxides, UO$_{2}$ and NpO$_{2}$ are two striking examples. UO$_{2}$ realizes a transverse 3$q$ magnetic dipolar and a $3q$ electric quadrupolar order~\cite{PhysRevLett.105.167201,PhysRevB.40.1856}, while NpO$_{2}$ is characterized by a high-rank magnetic multipolar order~\cite{PhysRevB.78.104425,RevModPhys.81.807}. In the late members of the actinide dioxides, PuO$_{2}$ is non-magnetic due to its $5f^{4.0}$ electronic configuration~\cite{Yasuoka901}. AmO$_{2}$ is supposed to have a longitudinal 3$q$ high-rank multipolar ordered state~\cite{PhysRevB.88.195146}, similar to NpO$_{2}$. The magnetic behaviors in these actinide compounds can be more or less explained theoretically. However, the magnetic properties of CmO$_{2}$ are totally unexpected and remain unsolved up to now~\cite{wen:2013}.

Based on the ionic picture, it is generally accepted that the valences of actinide ions in actinide dioxides are tetravalent. We thus naively reckon that the Cm ions in CmO$_{2}$ obey this rule as well. Note that there are six 5$f$ electrons for Cm$^{4+}$ ion. According to the Hund's rules, the ground state of Cm$^{4+}$ ion should be a singlet with $J = 0$, $S = 3$, and $L = 3$. Undoubtedly, it is a non-magnetic state. However, the temperature dependence of the magnetic susceptibility demonstrated a large effective moment of 3.4~$\mu_{\text{B}}$, and the analysis of the neutron diffraction pattern established that the CmO$_{2}$ sample was no long-range magnetic order~\cite{MORSS1989273}. As first glance, the non-magnetic ground state is inconsistent with the experimental results. In order to resolve this contradiction, several mechanisms have been proposed. A straightforward explanation is to consider the effect of Cm$^{3+}$ magnetic impurity, which owns seven 5$f$ electrons~\cite{PhysRevB.75.115107}. Its ground state multiplet is characterized by $J = 7/2$, $S = 7/2$, and $L = 0$. However, this explanation is already excluded by previous experiments. First, the samples are proven to be very close to stoichiometry. Second, if there is a mixture of Cm$^{3+}$ and Cm$^{4+}$, the oxygen sub-lattice should be rearranged. But there is no experimental evidence for the corresponding super-lattice peak in neutron diffraction spectra~\cite{MORSS1989273}. Another possibility to understand this problem is to consider the magnetic excited state of the Cm$^{4+}$ ions~\cite{PhysRevB.28.2317}. Due to the interplay of Coulomb interactions, spin-orbit coupling, and crystalline electric field, the excitation energy is likely to be smaller than the value of Land\'{e} internal rule. Consequently, Niikura and Hotta suggested that with carefully chosen parameters, the magnetic behaviors of CmO$_{2}$ can be well reproduced by solving an Anderson impurity model~\cite{PhysRevB.83.172402}. They concluded that the effective magnetic moment should reduce with decrease of the temperature. Once the temperature is low enough the magnetic behaviors should disappear. This mechanism has not been confirmed by experiments as well. In addition to the two proposals, Prodan \emph{et al.} suggested a covalent picture for CmO$_{2}$. In this picture, Cm could borrow electrons from O-$2p$ orbitals to achieve the stable half-filled $5f$-shell configuration~\cite{PhysRevB.76.033101}. They have performed screened hybrid functional calculations. But the calculated lattice parameter deviates considerably from the measured value and the obtained 5$f$ occupancy is only 6.2, which is too small to support the experimental magnetic moment.       

Besides these puzzling magnetic behaviors, CmO$_{2}$ has already distinguished itself from the other actinide dioxides for its intriguing ground state properties. For example, the lattice constants for $An$O$_{2}$ decrease monotonically with increasing atomic number~\cite{morss_book}. However, the lattice constants in CmO$_{2}$ deviate from this general trend obviously~\cite{asprey:1955,PETERSON19714111}. A cusp is noticed at CmO$_{2}$ in the plot of the $An$O$_{2}$ lattice constants $a_{0}$ vs atomic number $Z$~\cite{PhysRevB.76.033101}. Another example is about the single-particle properties of CmO$_{2}$. The actinide dioxides are usually charge-transfer insulators or Mott-Hubbard insulators with large band gaps ($E_{\text{gap}} >$ 1~eV)~\cite{wen:2013}. However, the band gap of CmO$_{2}$ has not been determined experimentally, but most of theoretical calculations have predicted a very small band gap ($E_{\text{gap}} \sim$ 0.4~eV)~\cite{PhysRevB.76.033101,PhysRevB.96.235137}.   

Apparently, not only the magnetic behaviors, but also the ground state electronic structure of CmO$_{2}$ remains mysterious. In the present work, we would like to address these problems, and propose a new mechanism to explain its magnetic puzzle. We endeavored to study the ground state electronic structure of CmO$_{2}$ by means of a state-of-the-art first-principles many-body approach. Our results imply that the oxidation state of Cm ions in CmO$_{2}$ is probably non-integer. In other words, it is neither Cm$^{3+}$ [i.e., Cm (III)], nor Cm$^{4+}$ [i.e., Cm (IV)]. Instead, it is mixed-valence with heavy 5$f$ valence state fluctuation. The 5$f$ electrons fluctuate among various electronic configurations (mainly including 5$f^{6.0}$ and 5$f^{7.0}$), and thus lead to a sizable magnetic moment.

The rest of manuscript is organized as follows. In Sec.~\ref{sec:method}, the methodology and computational details are introduced briefly. Section~\ref{sec:results} is the major part of this manuscript. All of the calculated results are presented here. In Sec.~\ref{sec:discuss}, we make a detailed comparison about the available mechanisms which could explain CmO$_{2}$'s magnetic moment. In addition, we further discuss the influence of temperature and pressure effects on the electronic structure of CmO$_{2}$. Finally, Sec.~\ref{sec:summary} provides a short summary. 


\section{method\label{sec:method}}

\begin{figure}[t!]
\centering
\includegraphics[width=\columnwidth]{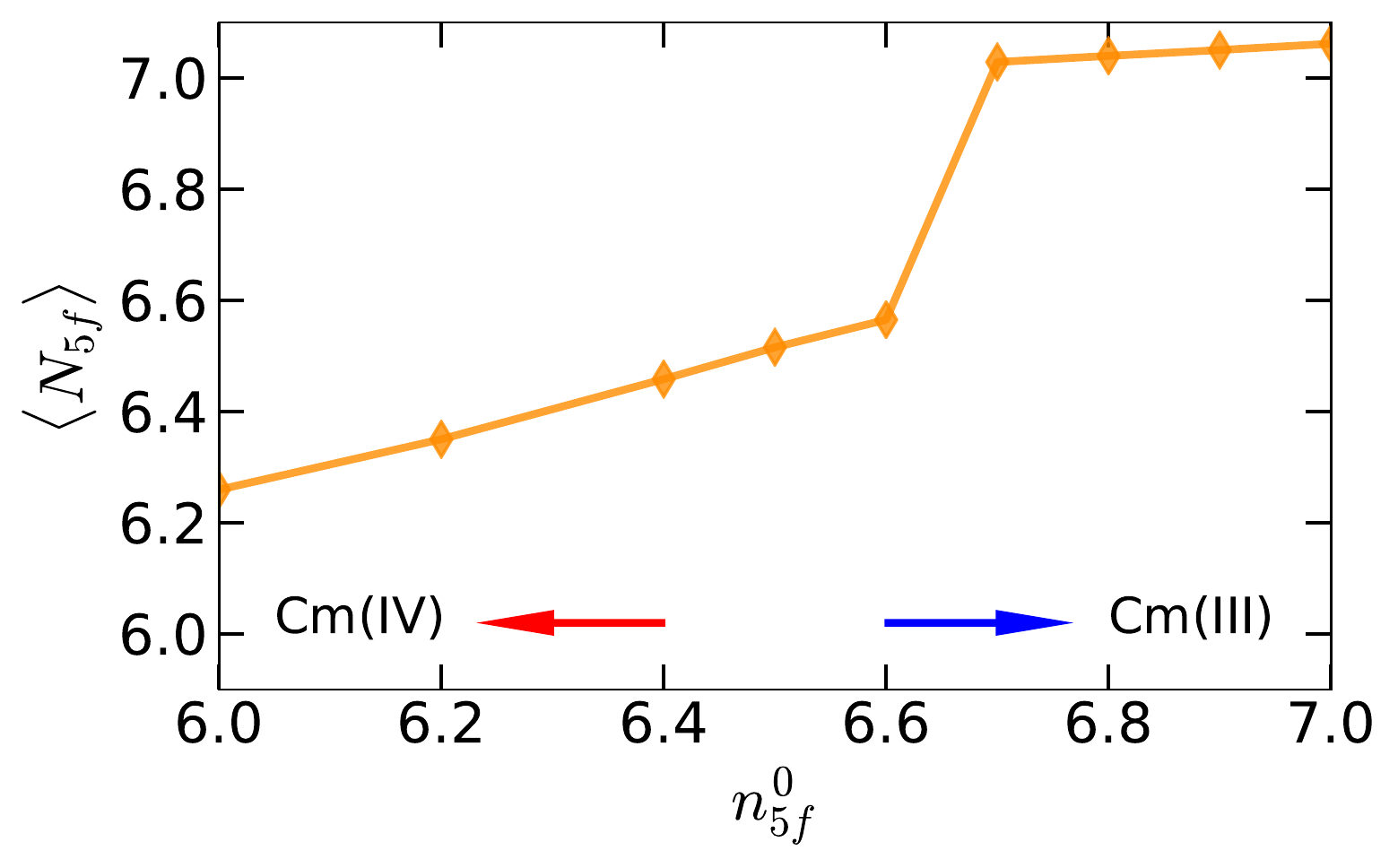}
\caption{(Color online). Total 5$f$ occupancy $\langle N_{5f} \rangle$ as a function of nominal 5$f$ occupancy $n^{0}_{5f}$. $\langle N_{5f} \rangle$ is measured by the CT-HYB impurity solver~\cite{RevModPhys.83.349}. See text for more details. \label{fig:n5f}}
\end{figure}

In 5$f$ electronic systems, both Coulomb interactions and spin-orbit coupling play vital roles~\cite{RevModPhys.81.235}. Therefore, it is essential to take them into considerations at the same footing~\cite{shim:2007}. In the present work, we employed the dynamical mean-field theory, in combination with the density functional theory (dubbed as DFT + DMFT) to accomplish this job~\cite{RevModPhys.78.865,RevModPhys.68.13}.

We utilized the \texttt{WIEN2K} software package~\cite{wien2k}, which implements a full-potential linearized augmented plane-wave formalism, to carry out the band structure calculations. The experimental crystal structure of CmO$_{2}$ was used~\cite{asprey:1955,PETERSON19714111,KONINGS2001255}. The generalized gradient approximation, specifically the Perdew-Burke-Ernzerhof functional~\cite{PhysRevLett.77.3865}, was selected to describe the exchange-correlation potential. The spin-orbit coupling was included. The $K$-mesh for Brillouin zone integration was 17 $\times $ 17 $\times $ 17, and $R_{\text{MT}} K_{\text{MAX}} = 8.0$. The muffin-tin radii for Cm and O ions were 2.5~au and 1.7~au, respectively.

The strong correlated nature of Cm's 5$f$ electrons were treated in a non-perturbed manner by DMFT. We adopted the \texttt{eDMFT} software package, which was implemented by K. Haule \emph{et al.}~\cite{PhysRevB.81.195107} The Coulomb interaction matrix for Cm's 5$f$ electrons were parameterized by applying the Slater integrals $F^{(k)}$. The Coulomb repulsion interaction parameter $U$ and Hund's exchange interaction parameter $J_{\text{H}}$ were 7.5~eV and 0.6~eV, respectively~\cite{RevModPhys.81.235}. The double counting term for 5$f$ self-energy functions was subtracted via the fully localized limit scheme~\cite{jpcm:1997}. The expression reads:
\begin{equation}
\label{eq:dc}
\Sigma_{\text{dc}} = U \left(n^{0}_{5f} - \frac{1}{2}\right) - \frac{J_{\text{H}}}{2} \left(n^{0}_{5f} - 1.0\right), 
\end{equation}
where $n^{0}_{5f}$ denotes the nominal 5$f$ occupation number. In the present works, we tried various $n^{0}_{5f}$ ($n^{0}_{5f} \in [6.0,7.0]$) to mimic the different oxidation states of Cm ions (see Fig.~\ref{fig:n5f}). For instance, for Cm~(IV), $n^{0}_{5f} = 6.0$, while for Cm~(III), $n^{0}_{5f} = 7.0$~\cite{t1}. Notice that $n^{0}_{5f}$ is just an artificial parameter, it shouldn't be altered during the calculations~\cite{t2}. The hybridization expansion continuous-time quantum impurity solver (dubbed as CT-HYB)~\cite{PhysRevLett.97.076405,PhysRevB.75.155113,RevModPhys.83.349} was employed to solve the resulting 14-bands Anderson impurity models. The number of quantum Monte Carlo sweeps was $2 \times 10^{8}$ per CPU process. The system temperature was approximately 116~K ($\beta = 1/T = 100.0$), and the system was restricted to be paramagnetic~\cite{t3}. We performed fully charge self-consistent DFT + DMFT calculations. The number of DFT + DMFT iterations was about $60 \sim 80$.


\begin{table*}[htbp]
\caption{Bulk properties, including equilibrium lattice constants $a_0$ (\AA), bulk modulus $B$ (GPa) and its first derivative with respect to the pressure ($B^{\prime}$), and band gap $E_{\text{gap}}$ (eV) of CmO$_{2}$. \label{tab:bulk}}
\begin{ruledtabular}
\begin{tabular}{rccccccccc}
                 & DFT + DMFT\footnotemark[1] 
                 & DFT + DMFT\footnotemark[2] 
                 & DFT + $U$ + SOC\footnotemark[3] 
                 & DFT + $U$ + SOC\footnotemark[4]
                 & DFT + $U$ + SOC\footnotemark[5]
                 & HSE\footnotemark[6]
                 & HSE + SOC\footnotemark[6] 
                 & SIC\footnotemark[7] 
                 & Exp.\footnotemark[8] \\
\hline
$a_0$            & 5.45   & 5.42   &  5.410  &  5.523    &  5.490   & 5.365    & 5.360  & 5.37   & 5.359 \\
$B$              & 223.1  & 235.5  &  193.3  &  129.6    &  123.7   &          &        & 212.0  & 218.0 \\
$B^{\prime}$     & 4.35   & 1.52   &  4.4    &    4.7    & 5.6      &          &        &        & 7.0   \\
$E_{\text{gap}}$ & 3.5    & 1.8    &  1.94   &  metallic & metallic & 0.4      & 0.4    &  0.4   &\\
\end{tabular}
\end{ruledtabular}
\footnotetext[1]{The present work. Cm (III) case.}
\footnotetext[2]{The present work. Cm (IV) case.}
\footnotetext[3]{For non-magnetic state of CmO$_{2}$. See Ref.~[\onlinecite{PhysRevB.96.235137}].}
\footnotetext[4]{For ferromagnetic state of CmO$_{2}$. See Ref.~[\onlinecite{PhysRevB.96.235137}].}
\footnotetext[5]{For antiferromagnetic state of CmO$_{2}$. See Ref.~[\onlinecite{PhysRevB.96.235137}].}
\footnotetext[6]{For antiferromagnetic state of CmO$_{2}$. See Ref.~[\onlinecite{wen:2013}].}
\footnotetext[7]{For antiferromagnetic state of CmO$_{2}$. See Ref.~[\onlinecite{PhysRevB.81.045108}].}
\footnotetext[8]{See Ref.~[\onlinecite{KONINGS2001255}] and [\onlinecite{dan2002}].}
\end{table*}

\begin{figure*}[ht]
\centering
\includegraphics[width=\textwidth]{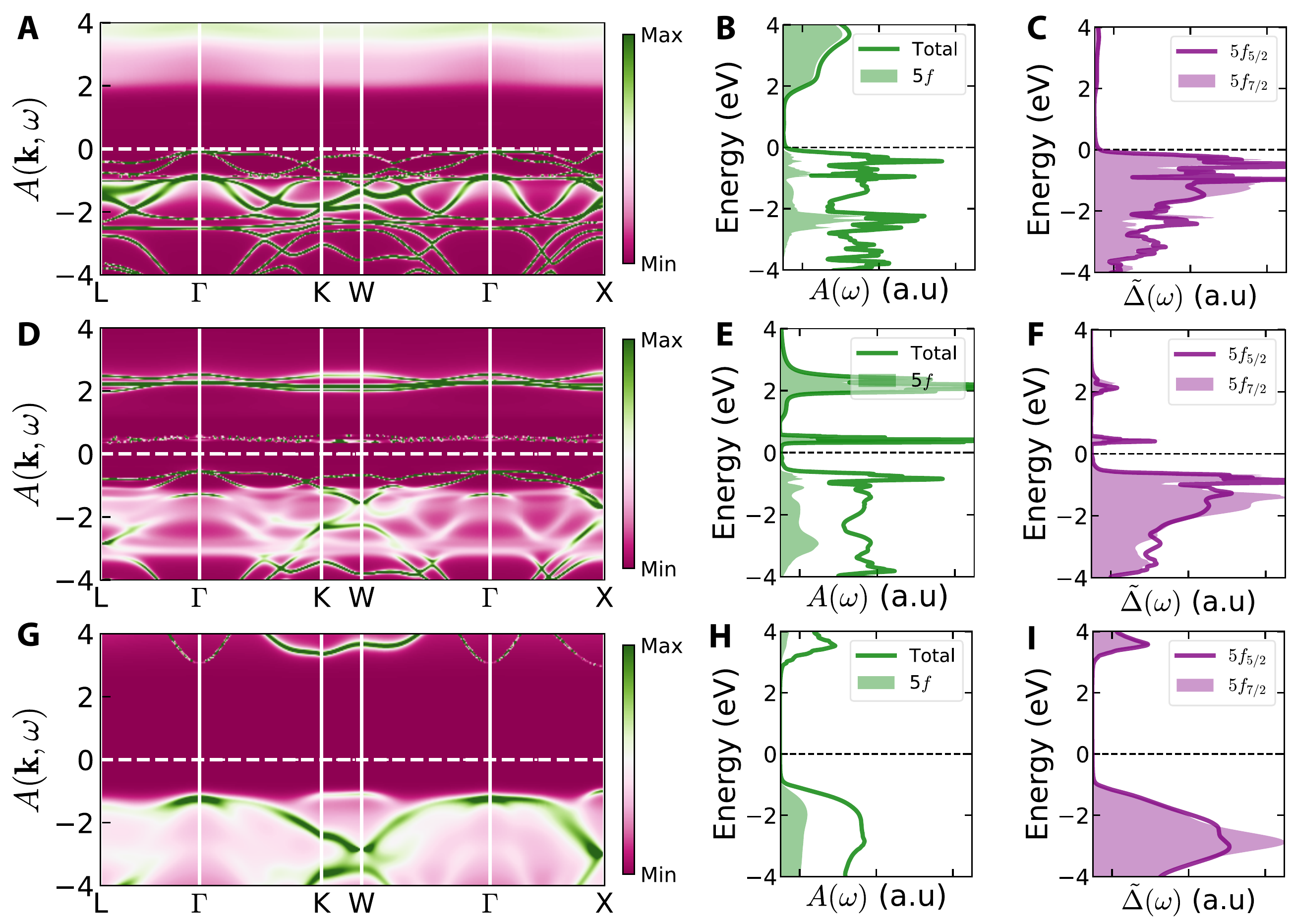}
\caption{(Color online). Ground state electronic structure of CmO$_{2}$ obtained by the DFT + DMFT method. The results for the Cm (IV) configuration ($n^{0}_{5f} = 6.0$) are shown in panels (a)-(c), while those for the Cm (III) configuration ($n^{0}_{5f} = 7.0$) are shown in panels (g)-(i). The results shown in panels (d)-(f) are for the intermediate configuration ($n^{0}_{5f} = 6.5$). (a), (d), and (g) Momentum-resolved spectral functions $A(\mathbf{k},\omega)$. (b), (e), and (h) Total density of states (thick solid lines) and 5$f$ partial density of states (colored shadow areas). (c), (f), and (i) Imaginary parts of hybridization functions. The $5f_{5/2}$ and $5f_{7/2}$ manifolds are represented by solid lines and colored shadow areas, respectively. The horizontal dashed lines denote the Fermi level. \label{fig:akw}}
\end{figure*}

\begin{figure*}[ht]
\centering
\includegraphics[width=\textwidth]{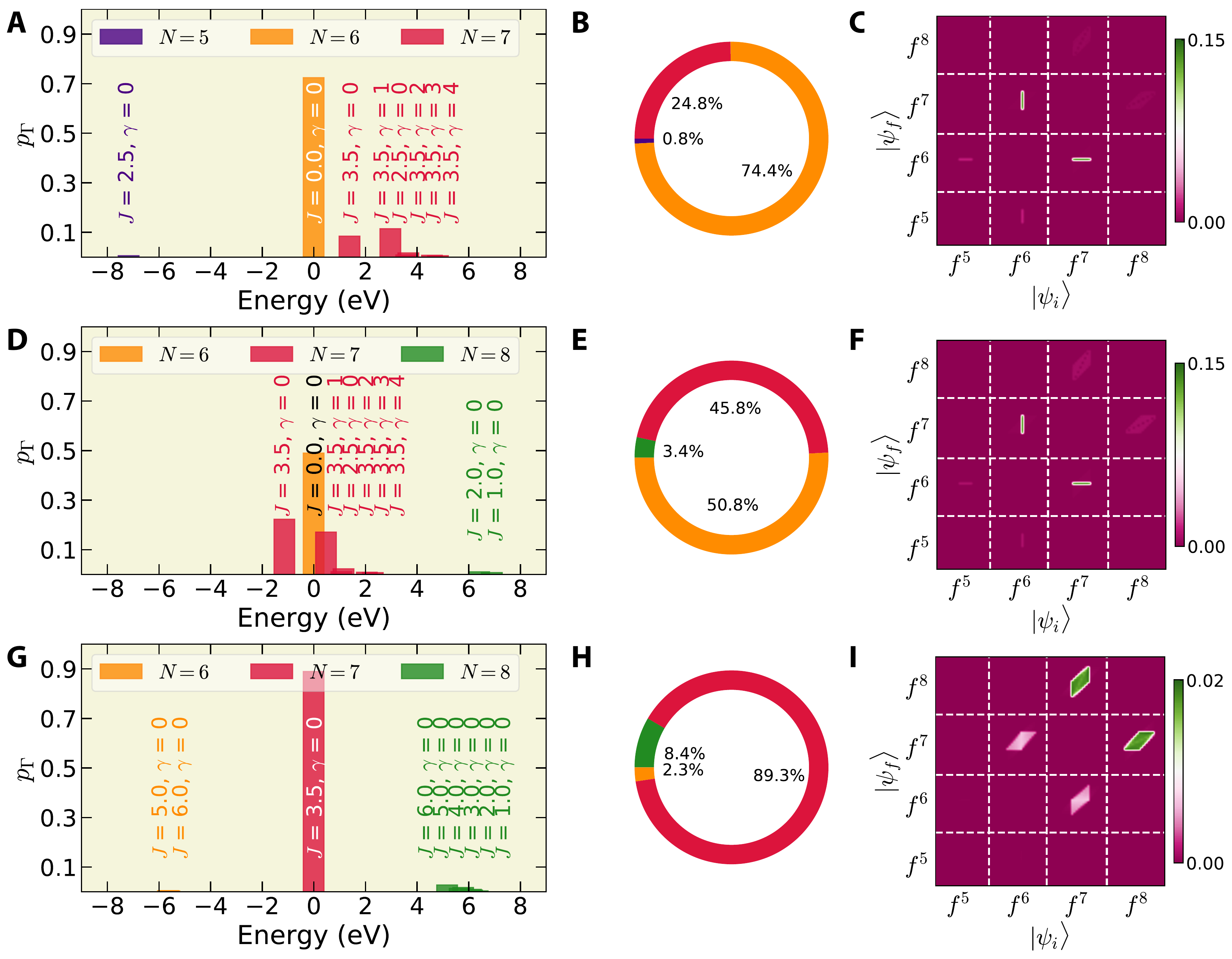}
\caption{(Color online). Valence state fluctuations in CmO$_{2}$ obtained by the DFT + DMFT method. The results for the Cm (IV) configuration ($n^{0}_{5f} = 6.0$) are shown in panels (a)-(c), while those for the Cm (III) configuration ($n^{0}_{5f} = 7.0$) are shown in panels (g)-(i). The results shown in panels (d)-(f) are for the intermediate configuration ($n^{0}_{5f} = 6.5$). (a), (d), and (g) Valence state histogram (or equivalently atomic eigenstates probability). Here the atomic eigenstates are labelled by using three good quantum numbers, namely $N$ (total occupancy), $J$ (total angular momentum), and $\gamma$ ($\gamma$ stands combination of the rest of good quantum numbers). Most of the atomic eigenstates with trivial contributions are not shown in these panels. (b), (e), and (h) Distribution of atomic eigenstates with respect to total occupancy $N$. (c), (f), and (i) Transition probabilities between any two atomic eigenstates. Here $|\psi_i\rangle$ and $|\psi_f\rangle$ denote the initial and final states, respectively. \label{fig:prob}}
\end{figure*}

\begin{figure}[ht]
\centering
\includegraphics[width=\columnwidth]{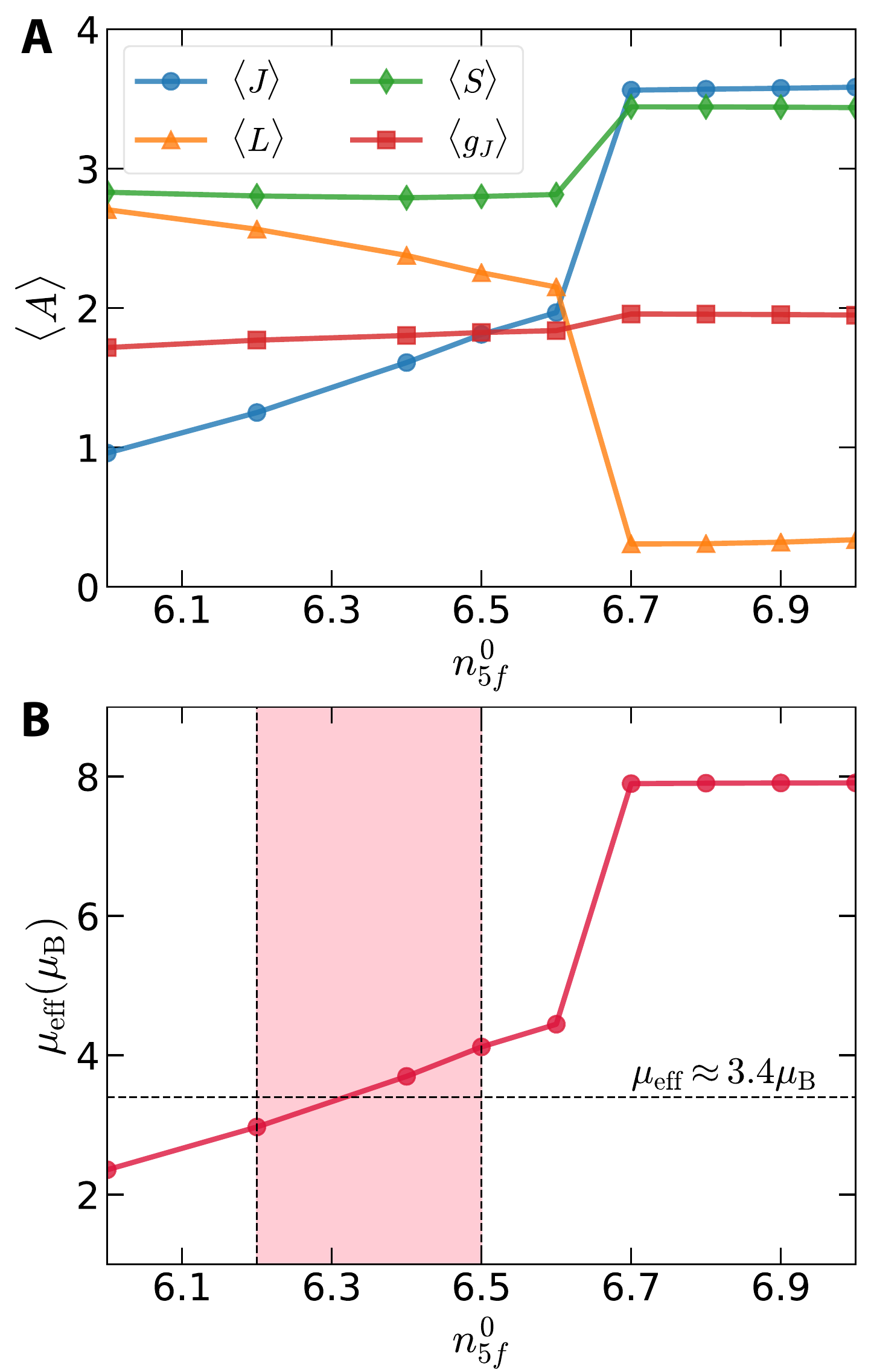}
\caption{(Color online). (a) Effective total angular momentum $\langle J \rangle$, orbital momentum $\langle L \rangle$, spin momentum $\langle S \rangle$, and Land\'{e} $g$ factor $\langle g_J \rangle$ as a function of $n^{0}_{5f}$. (b) Effective local magnetic momentum $\mu_{\text{eff}}$ as a function of $n^{0}_{5f}$. The experimental value, which is denoted by a horizontal dashed line, is taken from Ref.~[\onlinecite{MORSS1989273}]. Here the pink region marks the most likely oxidation oxides of Cm ions ($n^{0}_{5f} \in [6.2,6.5]$). Their local magnetic moments are plausible and close to the experimental value. \label{fig:gj}}
\end{figure}

\begin{figure}[ht]
\centering
\includegraphics[width=\columnwidth]{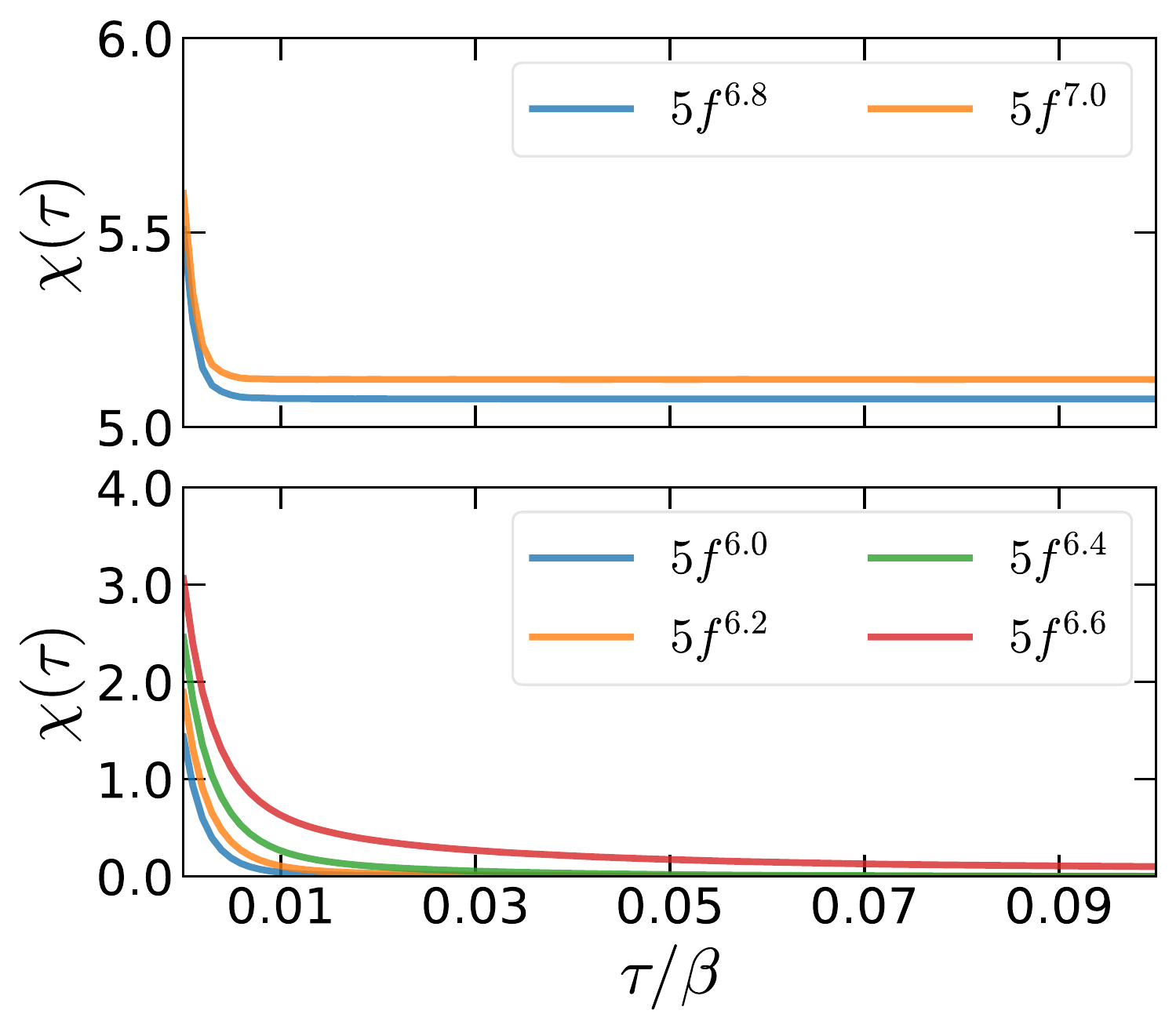}
\caption{(Color online). Imaginary-time spin-spin correlation functions of CmO$_{2}$, $\chi(\tau) = \langle S_z (\tau) S_z (0) \rangle$. (Top) For low oxidation states of Cm ions ($n^{0}_{5f} =$ 6.8 and 7.0). (Bottom) For high oxidation states of Cm ions ($n^{0}_{5f} =$ 6.0, 6.2, 6.4, and 6.6). \label{fig:chi}}
\end{figure}

\section{results\label{sec:results}}

\subsection{Bulk properties}

Previous attempts to apply the \emph{ab initio} methods to study the actinide-based materials have been hampered by the lack of highly accurate and efficient first-principles approaches that can correctly capture and describe the correlated nature of $5f$ electrons. Though DFT + DMFT might be one of the most powerful approaches that ever established to study the strongly correlated electron materials~\cite{RevModPhys.68.13,RevModPhys.78.865}, it has been seldom applied to the late actinides and their compounds~\cite{PhysRevB.99.045109,shim:2007,PhysRevLett.96.036404}. Here, in order to build confidence in the DFT + DMFT approach, we tried to calculate the equilibrium properties, including lattice constants $a_{0}$ and bulk modulus $B$ of CmO$_{2}$. We at first calculated the $E-V$ curve, and then used the Birch-Murnaghan equation of states~\cite{PhysRev.71.809} to fit it. Our results, together with the other theoretical and experimental values where available are summarized in Table.~\ref{tab:bulk}.

It is clear that our results agree closely with the experimental and the other theoretical values. For the Cm (III) case, the deviations are only $\sim$ 1.7\% for $a_0$ and $\sim$ 2.3\% for $B$, respectively. As for the Cm (IV) case and the other potential oxidation states (they are not shown in this table), the deviations are similar or even smaller. These facts suggest that the bulk properties of CmO$_{2}$ are very well reproduced by the DFT + DMFT approach, irrespective of the oxidation states of Cm ions. Actually, Cm ions were assumed to be Cm (IV) in most of the previous theoretical calculations~\cite{PhysRevB.81.045108,wen:2013}.

\subsection{Quasiparticle band structures}

The quasiparticle band structures or momentum-resolved spectral functions $A(\mathbf{k},\omega)$ of CmO$_{2}$ within various oxidation states [Cm~(IV), intermediate configuration ($n^{0}_{5f} = 6.5$), and Cm~(III)] are depicted in Fig.~\ref{fig:akw}(a), (d), and (g). The following characteristics are noticeable: (i) All figures show considerable band gaps (see Table~\ref{tab:bulk}). Note that the band gap of the Cm (III) case is much larger than those of the other cases. It is not surprised because when the correlated orbital is in half-filling, it would suffer larger effective interaction. This rule has been revealed in strongly correlated $d$-electron systems and model Hamiltonian calculations~\cite{PhysRevLett.107.256401}. (ii) The band gaps of the Cm (IV) case and the intermediate configuration are indirect ($\Gamma \to K$), which are consistent with the previous theoretical results~\cite{PEGG2017269,PhysRevB.88.195146,PhysRevB.96.235137}. Interestingly, the band gap of the Cm (III) case is direct. (iii) We observe stripe-like patterns in these figures. These features exist at approximately -3~eV $< \omega <$ -1~eV and $\omega > 2$~eV for the Cm (IV) case, -4~eV $< \omega <$ -1~eV for the intermediate configuration, and $\omega < -1.5$~eV for the Cm (III) case. (iv) In addition, we also observe narrow and flat bands in the vicinity of 0.4~eV and 2.0~eV for the intermediate configuration. The stripe-like patterns and flat bands are mainly from the contributions of localized 5$f$ electrons. These 5$f$ bands look quite blurred, which indicate that the 5$f$ electrons are incoherent.

From the total density of states $A(\omega)$ and $5f$ partial density of states $A_{5f}(\omega)$, we can see that the band gaps are associated not only with the transitions between occupied and unoccupied 5$f$ states, but also with the transitions between 5$f$ and the other weakly correlated or non-correlated orbitals [see Fig.~\ref{fig:akw}(b), (e), and (h)]. According to the results of the three representative configurations, we can establish that CmO$_{2}$ is a typical charge transfer insulator with a sizable band gap ($>$ 1~eV), instead of Mott-Hubbard insulator. Previous theoretical studies predicted that CmO$_{2}$ is a charge transfer insulator, but with a small band gap ($< 1$ eV), or even a semi-metal~\cite{PhysRevB.88.195146,PhysRevB.96.235137,wen:2013,PhysRevB.81.045108}. To validate these predictions, further photoemission or optical experiments are highly desired~\cite{PhysRevB.15.2929}.

From Fig.~\ref{fig:akw}(a), there exist significant overlaps between the 5$f$ bands and $spd$ conduction bands between -3~eV and -1~eV. It means that there must be strong $c-f$ hybridization for the Cm (IV) case. For the intermediate configuration, strong $c-f$ hybridization is seen between -4~eV and -1~eV [see Fig.~\ref{fig:akw}(d)]. For the Cm (III) case, similar phenomenon is observed when $\omega < 1.5$~eV [see Fig.~\ref{fig:akw}(g)]. Note that the 5$f$ hybridization function $\Delta(\omega)$ is an ideal measurement for the $c-f$ hybridization effect. In Fig.~\ref{fig:akw}(c), (f), and (i), the 5$f$ hybridization functions are shown. Here, we only depict the imaginary parts, i.e., $\tilde{\Delta}(\omega) = - \text{Im} \Delta(\omega) / \pi$. Due to the spin-orbit coupling effect, the hybridization functions are split into two parts, $5f_{5/2}$ and $5f_{7/2}$ components. We can see that there are exactly strong hybridizations between the $5f$ bands and the other bands below the Fermi level. The $c-f$ hybridization in the Cm~(III) case is slightly weaker than the others. Strong $c-f$ hybridization enables the valence electrons to transfer between 5$f$ and $spd$ bands, and finally leads to the so-called mixed-valence or valence state fluctuation behavior as discussed below.  

\subsection{Valence state fluctuations}

Valence state fluctuation is an ubiquitous feature in strongly correlated $f$-electron systems~\cite{PhysRevB.81.035105,shim:2007,Janoscheke:2015}. It is very sensitive to external conditions, such as temperature, pressure, alloying, etc~\cite{PhysRevB.94.075132,PhysRevB.98.195102,PhysRevB.99.045109,PhysRevB.99.045122}. Naturally, we expected that valence state fluctuation behaviors should be quite different for the Cm (III), intermediate configuration, and Cm (IV) cases. Valence state histogram may be the most suitable quantity to qualify the valence state fluctuation. It denotes the probability to find out a valence electron in a given atomic eigenstate $|\psi_{\Gamma}\rangle$, which is labelled by using some good quantum numbers (such as total occupancy $N$ and total angular momentum $J$)~\cite{PhysRevB.75.155113,shim:2007}. Thus, we tried to calculate this physical quantity via the CT-HYB quantum impurity solver~\cite{RevModPhys.83.349,PhysRevLett.97.076405}. The calculated results are illustrated in Fig.~\ref{fig:prob}. (i) For the Cm (IV) case, the valence state fluctuation is rather weak [see Fig.~\ref{fig:prob}(a)]. The predominant atomic eigenstate is $| N = 6.0, J = 0.0, \gamma = 0.0 \rangle$. It accounts for approximately 70\%. The residual contributions are mainly from the atomic eigenstates with $N = 7$. Though the contributions from the other high-lying atomic eigenstates are quite small, they are not trivial and have to be taken into considerations explicitly. (ii) As for the Cm (III) case, it manifests the weakest valence state fluctuation [see Fig.~\ref{fig:prob}(g)]. The atomic eigenstate $| N = 7.0, J = 3.5, \gamma = 0.0 \rangle$ become overwhelming. It accounts for approximately 90\%, which means the $5f$ electrons are virtually locked at this atomic eigenstate (the ground state). The contributions from the other atomic eigenstates are indeed trivial. (iii) For the intermediate configuration, we observe heavy 5$f$ valence state fluctuation [see Fig.~\ref{fig:prob}(d)]. There are at least three atomic eigenstates ($|N = 6, J = 0, \gamma = 0\rangle$, $|N = 7, J = 3.5, \gamma = 0\rangle$, and $|N = 7, J = 3.5, \gamma = 1\rangle$), whose contributions are comparable. (iv) Finally, from the perspective of valence state fluctuation, we have intermediate configuration $ > $ Cm~(IV) $ > $  Cm~(III). The strong $c-f$ hybridization may be the driving force of $5f$ valence state fluctuation.

The distributions of valence electron configurations are summed up and shown in Fig.~\ref{fig:prob}. Clearly, these systems are mixed-valence. For the Cm (IV) case, it is a mixture of $5f^{5.0}$, $5f^{6.0}$, and $5f^{7.0}$ electronic configurations [see Fig.~\ref{fig:prob}(b)]. As a consequence, its effective 5$f$ occupancy $\langle N_{5f} \rangle$ is about 6.26. As for the Cm (III) case, its ground state comprises $5f^{6.0}$, $5f^{7.0}$, and $5f^{8.0}$ electronic configurations [see Fig.~\ref{fig:prob}(h)], so $\langle N _{5f} \rangle \approx 7.06$. Similar, we obtain that $\langle N _{5f} \rangle \approx 6.52$ for the intermediate configuration [see Fig.~\ref{fig:prob}(e)]. The relationship between $n^{0}_{5f}$ and $ \langle N_{5f} \rangle $ has been shown in Fig.~\ref{fig:n5f}.

As a byproduct, we can use the information of valence state fluctuation to make a rough estimation about the effective magnetic moment. Our strategy is as follows. First of all, we use the following equation to evaluate the expected value of a given quantum mechanical operator $A$: 
\begin{equation}
\langle A \rangle = \sum_{\Gamma} p_{\Gamma} A_{\Gamma}, 
\end{equation} 
where $p_{\Gamma}$ is the probability of any atomic eigenstate $|\psi_{\Gamma}\rangle$, and $\langle A\rangle$ could be $\langle N_{5f} \rangle$, $\langle J\rangle $, $\langle L\rangle $, and $\langle S \rangle$. Next, we can calculate the Land\'{e} $g$ factor $g_J$ through:
\begin{equation}
\langle g_J \rangle = \frac{3}{2} + 
\frac{\langle S \rangle (\langle S \rangle +1) - \langle L\rangle (\langle L \rangle +1)}
{2\langle J \rangle (\langle J \rangle +1)}. 
\end{equation}
Finally, the effective magnetic momentum $\mu_{\text{eff}}$ can be calculated via the following relation~\cite{PhysRevB.83.125111}:
\begin{equation}
\mu_{\text{eff}} \approx \langle g_{J} \rangle \sqrt{\langle J \rangle (\langle J \rangle +1)}.
\end{equation}
The calculated results are summarized in Fig.~\ref{fig:gj}.

For the Cm (IV) case, $\langle J \rangle \approx 0.96$, $\langle L \rangle \approx 2.71$, and $\langle S \rangle \approx 2.83$. For the Cm (III) case, $\langle J \rangle \approx 3.59$, $\langle L \rangle \approx 0.34$, and $\langle S \rangle \approx 3.44$. All these values deviate apparently from the values at atomic limits. In case $n^{0}_{5f}$ goes from 6.0 to 7.0, we can split this process into three distinct stages according to the changes in $\langle J \rangle$ and $\mu_{\text{eff}}$. (i) When $n^{0}_{5f} < 6.6$, $\langle J \rangle$ and $\mu_{\text{eff}}$ increase monotonously with respect to $n^{0}_{5f}$. Further, we find that $\langle J \rangle-n^{0}_{5f}$ and $\mu_{\text{eff}}-n^{0}_{5f}$ exhibit quasi-linear relations. (ii) When $6.6 < n^{0}_{5f} < 6.7$, sudden jumps for $\langle J \rangle$ and $\mu_{\text{eff}}$ are discerned. It is speculated that a magnetic or electronic transition would occur here. (iii) When $n^{0}_{5f} > 6.7$, $\langle J \rangle$ and $\mu_{\text{eff}}$ approach their saturated values, 3.60 and 8.0~$\mu_{\text{B}}$, respectively. Notice that the Land\'{e} $g$ factor almost remain constants during this process ($\langle g_J \rangle \approx 2.0$). From the calculated and experimental values of $\mu_{\text{eff}}$, we can conclude that the oxidation state of Cm ions in CmO$_{2}$ is neither Cm (III), nor Cm (IV)~\cite{PhysRevB.28.2317,MORSS1989273}. It should be an intermediate configuration. In Fig.~\ref{fig:gj}(b), the possible oxidation states are highlighted by pink region. We have $6.2 < n^{0}_{5f} < 6.5$ and the corresponding 5$f$ occupancy $6.35 < \langle N_{5f} \rangle < 6.52$ (see Fig.~\ref{fig:n5f}).    

In Fig.~\ref{fig:prob}, we also plot the transition probabilities between any two atomic eigenstates, $\Pi(\langle \psi_f | \psi_i\rangle)$. Here, $| \psi_i \rangle$ and $| \psi_f \rangle$ denote the initial and final states, respectively. We can see intense many-body transitions between $5f^{6.0}$ and $5f^{7.0}$ states for the Cm (IV) case [see Fig.~\ref{fig:prob}(c)] and the intermediate configuration [see Fig.~\ref{fig:prob}(f)]. For the Cm (III) case, the transitions between $5f^{7.0}$ and $5f^{8.0}$ states become dominant [see Fig.~\ref{fig:prob}(i)]. Overall, the intermediate configuration exhibit stronger and more centralized many-body transitions. In the Cm (IV) case, though the magnetic excited states (i.e., the $5f^{7.0}$ states) exist, their contributions are not sufficient to reproduce the experimentally observed magnetic moment~\cite{MORSS1989273}.

\subsection{Spin-spin correlation functions}

To gain a deeper understanding about the magnetic properties of CmO$_{2}$, we further examine its imaginary-time spin-spin correlation functions $\chi(\tau)$, 
\begin{equation}
\chi(\tau) = \langle S_z (\tau) S_z (0) \rangle. 
\end{equation}
At first, we used the CT-HYB quantum impurity solver to measure the local dynamical susceptibility $\chi$ as a function of Matsubara frequencies $i\omega_n$~\cite{PhysRevB.75.155113,RevModPhys.83.349}. Later we converted it into imaginary-time axis to obtain $\chi(\tau)$, 
\begin{equation}
\chi(\tau) = \frac{1}{\beta} \sum_{n} e^{-i\omega_n \tau} \chi(i\omega_n). 
\end{equation}
The calculated results for $\chi(\tau)$ are illustrated in Fig.~\ref{fig:chi}. We find that the asymptotic behaviors of $\chi(\tau)$ for various $n^{0}_{5f}$ are totally different. (i) When $n^{0}_{5f} < 6.6$, $\chi(\tau)$ approaches zero very quickly. (ii) When $n^{0}_{5f} = 6.6$, $\chi(\tau)$ decreases slowly, and finally goes to a small finite value at large $\tau$. (iii) When $n^{0}_{5f} > 6.6$, $\chi(\tau)$ drops almost immediately, and then starts to keep a large fixed value at small $\tau$. These facts suggest that CmO$_{2}$ will fall into a spin-freezing-like phase~\cite{PhysRevLett.101.166405} once $n^{0}_{5f} \ge 6.6$. This conclusion is consistent with the evolution of $\mu_{\text{eff}}$ with respect to $n^{0}_{5f}$ [see Fig.~\ref{fig:gj}(b)]. 

Actually, we can use the following equation to evaluate the effective local magnetic moment $\mu_{\text{eff}}$ again~\cite{jk:2008}:
\begin{equation}
\mu_{\text{eff}} = \sqrt{T\chi_{\text{loc}}},
\end{equation}
where $\chi_{\text{loc}}$ is defined as 
\begin{equation}
\chi_{\text{loc}} = \int^{\beta}_0 d\tau \chi(\tau).
\end{equation}
Using the data presented in Fig.~\ref{fig:chi}, we reevaluated $\mu_{\text{eff}}$ as a function of $n^{0}_{5f}$. These values are smaller than those obtained via distributions of atomic eigenstates, but the overall trend is entirely similar. We confirm again that there will be a leap for $\mu_{\text{eff}}$ when $6.6 < n^{0}_{5f} < 6.7$. Only the intermediate configuration can yield a plausible local magnetic moment.  


\section{discussions\label{sec:discuss}}

\begin{figure}[t!]
\centering
\includegraphics[width=\columnwidth]{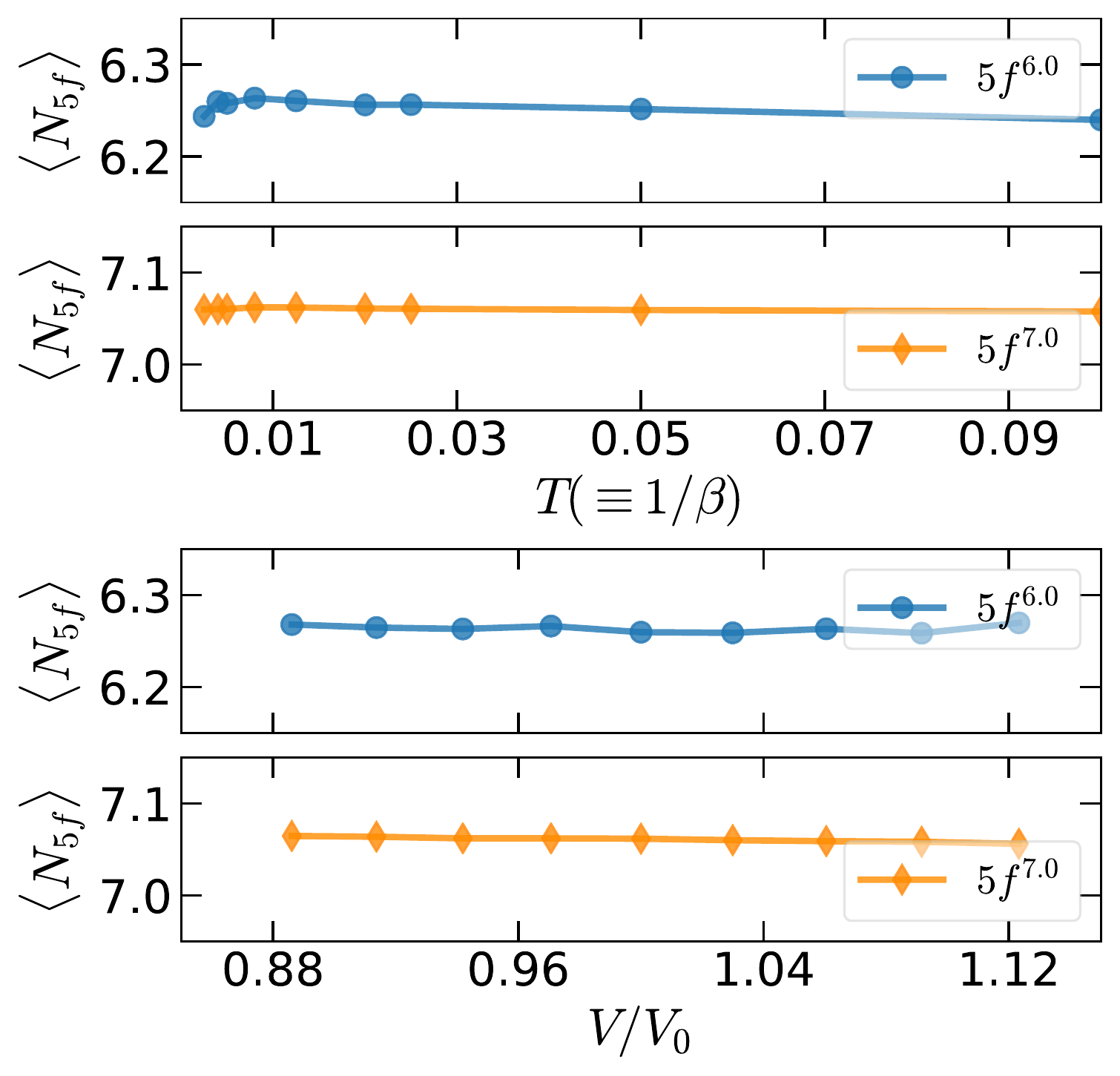}
\caption{(Color online). (Top) Temperature dependence of 5$f$ occupancies. (Bottom) Volume (or pressure) dependence of 5$f$ occupancies. Here $V_0 = 259.45$~bohr$^3$.~\label{fig:beta}}
\end{figure}

\emph{Puzzling magnetic behaviors in CmO$_{2}$.} As mentioned before, a few mechanisms have been proposed to explain the puzzling magnetic properties in CmO$_{2}$. Here we would like to summarize and compare their successes and failures. (i) \emph{Pure Cm~(IV) ions}. This explanation is incorrect because it yields a non-magnetic ground state. Even if we consider the effect of valence state fluctuation, the resulting magnetic moment is too small. Petit \emph{et al.} had employed the SIC-LSD method to study the electronic structures and magnetic properties of CmO$_{2}$ as well~\cite{PhysRevB.81.045108}. They assumed that the initial configuration of Cm ions is Cm~(IV). However, due to the spin-orbit coupling and hybridization, this initial configuration is significantly distorted and in fact the orbital moment is almost quenched (the spin moment is retained). Thus, a finite magnetic moment is obtained ($\mu_{\text{eff}} \approx 5.21$ $\mu_B$). This explanation sounds somewhat reasonable. But, the resulting 5$f$ occupancy is about 6.86, which is far from the $5f^{6.0}$ or $5f^{7.0}$ configuration. (ii) \emph{Pure Cm~(III) ions}. This explanation is excluded as well because it overestimates the magnetic moment. (iii) \emph{A mixture of Cm~(IV) and Cm~(III) ions}. This mechanism sounds reasonable. It can generate rational magnetic moment, but it requires rearrangement of oxygen sub-lattice. In other words, one way to satisfy Cm$^{3+}$ valence and O$^{2-}$ configuration at the same time would be to take the sesquioxide (Cm$_{2}$O$_{3}$), which is not supported by neutron diffraction experiment~\cite{MORSS1989273}. (iv) \emph{Cm~(IV) ions in excited states}. This mechanism is restricted by temperature. At low temperature region, it becomes invalid. (v) \emph{Covalent picture or intermediate configuration}. The spirit of this explanation is two-folds. At first, Cm can borrow additional electrons from O-$2p$ orbitals via the covalent bonds~\cite{PhysRevB.76.033101} or the mechanism of $c-f$ hybridization. Previous calculations using screened hybrid density functional theory revealed that the charge densities at oxygen sites deviate apparently from the expected values (O$^{2-}$)~\cite{PhysRevB.76.033101}. Actually, in our DFT + DMFT calculations, we find that $\langle N_{\text{O}_{2p}}\rangle$ is about 4.5 for $n^{0}_{5f} = 6.5$. It means that the O ions in CmO$_{2}$ probably do not form closed shells. Second, Cm's 5$f$ electrons could spend quite a lot of its lifetime in the atomic eigenstates with $N = 7$ and $N = 8$ due to the strong valence state fluctuation, leading to finite macroscopic moment. This mechanism do not break the cubic crystal symmetry. It is not sensitive to the change of temperature. More important, it is consistent with all the available experimental results~\cite{PhysRevB.75.115107,MORSS1989273,PhysRevB.28.2317,PhysRevB.15.2929}. Therefore, we believe that the intermediate configuration is more reasonable for the ground state electronic structure of CmO$_{2}$. 

\emph{Abnormal lattice constants in CmO$_{2}$.} The intermediate configuration also provides a possible explanation for the abnormal lattice constants in CmO$_{2}$. The formal valence of Cm ions in CmO$_{2}$ is +4. However, according to the present calculated results, its valence is non-integer and exhibits a significant trend toward +3. As for AmO$_{2}$ and the early members of $An$O$_{2}$, the formal expectations, i.e., $An^{4+}$ are well satisfied~\cite{PhysRevB.76.033101}. Therefore, it is easily to understand why the lattice constants of CmO$_{2}$ deviates the general trend and shows an evident cusp in the plot of $a_0-Z$ for $An$O$_{2}$. Actually, sesquioxides are more stable than dioxides for the late members of the actinide series. For actinides beyond Cf, only the sesquioxides have been observed experimentally~\cite{HAIRE1995185,MOORE1986187}. Clearly, CmO$_{2}$ sits at the boundary between $An^{3+}$ and $An^{4+}$.

\emph{Temperature- and pressure-dependent electronic structures.} We further examine the temperature dependence and pressure dependence of 5$f$ electronic configurations in CmO$_{2}$. The Cm (III) and Cm (IV) cases are taken as two representative examples. We find that the two oxidation states are quite stable against temperature and volume compression. In Fig.~\ref{fig:beta}, the calculated $5f$ occupancies as functions of temperature and volume are shown. Clearly, $\langle N_{5f} \rangle$ remains almost constants if we take numerical fluctuations into considerations. If we further increase the pressure (or reduce the volume), what will happen? Notice that CmO$_{2}$ will undergo a structural phase transition when the pressure is between 30 and 40~GPa. CmO$_{2}$ will transform from the cubic phase (space group: Fcc) to the orthorhombic structure (space group: Pnma), with about 10\% volume collapse~\cite{dan2002}. Similar volume changes have been observed in the high-pressure phase transitions of the other actinide dioxides (such AmO$_{2}$ and UO$_{2}$)~\cite{PhysRevB.70.014113,Huang_2017,doi:10.1080/08957950212818}, which are likely linked to the 5$f$ localized-itinerant crossover. As for CmO$_{2}$, it is expected that the 5$f$ occupancy will decrease (or equivalently, the 5$f$'s valence will increase, Cm$^{3+}$ $\to$ intermediate valence $\to$ Cm$^{4+}$) under pressure, and the local magnetic moment should be suppressed as well. 


\section{conclusions\label{sec:summary}}

In summary, we reported a systematic study of the ground state electronic structure in CmO$_{2}$. We carried out fully charge self-consistent DFT + DMFT calculations and considered various possible oxidation states of Cm ions. Our major findings are as follows. First, the ground state electronic configuration of Cm ions is neither Cm (III), nor Cm (IV). It should be an intermediate configuration. This configuration is very stable over a wide range of temperature or pressure. Second, the intermediate configuration leads to a macroscopic local moment in the nominally non-magnetic CmO$_{2}$ via the valence state fluctuation mechanism. The obtained 5$f$ occupancy is about 6.35 $\sim$ 6.52. Third, we predict that CmO$_{2}$ is a wide-gap charge transfer insulator (indirect band gap, $E_{\text{gap}} >$ 1.0~eV). Further experiments to verify our predictions and proposals are highly demanded. The x-ray magnetic circular dichroism (XMCD) would be an ideal tool to detect the mixed-valence behaviors of Cm ions in CmO$_{2}$, because it is extremely sensitive to the $f^{6}$ and $5f^{7}$ configurations and only needs microgram quantities for the experiments~\cite{PhysRevLett.114.097203}. 

\begin{acknowledgments}
We thank Prof. Gerry Lander for fruitful discussion with him. This work was supported by the Natural Science Foundation of China (No.~11874329, 11934020, and 11704347), and the Science Challenge Project of China (No.~TZ2016004).
\end{acknowledgments}


\bibliography{cm}

\begin{thebibliography}{58}
\bibitem [{t1()}]{t1}%
  \BibitemOpen
  \href@noop {} {}\bibinfo {note} {Here, we establish the maps between Cm (IV)
  and $n^{0}_{5f} = 6.0$, and between Cm (III) and $n^{0}_{5f} = 7.0$. However,
  when $n^{0}_{5f} = 6.0$, the resulting 5$f$ occupancy $\langle N_{5f}
  \rangle$ is about 6.26, instead of 6.0 (see Fig.~\ref{fig:n5f}). The
  situation is similar for the $n^{0}_{5f} = 7.0$ case.}\BibitemShut {Stop}%
\bibitem [{t2()}]{t2}%
  \BibitemOpen
  \href@noop {} {}\bibinfo {note} {We have tried several schemes for the double
  counting term. Only the fully localized limit scheme works. In principle,
  $n^{0}_{5f}$ should be determined self-consistently during the DFT + DMFT
  iterations. However, it fails to reproduce the insulating nature of
  CmO$_{2}$. So we have to fix $n^0_{5f}$ during the iterations and use it as a
  parameter to tune the real 5$f$ occupation number.}\BibitemShut {Stop}%
\bibitem [{t3()}]{t3}%
  \BibitemOpen
  \href@noop {} {}\bibinfo {note} {The ordering of CmO$_{2}$ is not expected,
  and has not been found by susceptibility measurements, at least down to about
  3~K.}\BibitemShut {Stop}%
\end{thebibliography}

\end{document}